\documentclass[superscriptaddress,showpacs,twocolumn,prl]{revtex4}

\usepackage{amssymb}
\usepackage{graphicx}
\usepackage{dcolumn}
\usepackage{bm}

\begin{document}

\title{Finite width and local field corrections to spin Coulomb drag
in a quasi-two-dimensional electron gas}
\author{S. M. Badalyan}
\email{Samvel.Badalyan@physik.uni-r.de} 
\affiliation{Department of Radiophysics,
Yerevan State University, 1 A. Manoukian St., Yerevan, 375025
Armenia} \affiliation{Department of Physics, University of
Regensburg, 93040 Regensburg, Germany}
\author{C. S. Kim}
\affiliation{Department of Physics and Institute for Condensed
Matter Theory, Chonnam National University, Gwangju 500-757,
Korea}
\author{G. Vignale}
\affiliation{Department of Physics and Astronomy, University of
Missouri - Columbia, Missouri 65211, USA}
\date{\today}

\begin{abstract}
We study the spin Coulomb drag  in a quasi-two-dimensional
electron gas of finite transverse width, including local field
corrections beyond the random phase approximation (RPA). We find
that the finite transverse width of the  electron gas causes a
significant reduction of the spin Coulomb drag. This reduction,
however, is largely compensated by the enhancement coming from the
inclusion of  many-body local field effects beyond the RPA,
thereby restoring good agreement with the experimental
observations by C. P. Weber \textit{et al.}, Nature, \textbf{437},
1330 (2005).
\end{abstract}

\maketitle


Lately a number of spin-based devices have been designed and
studied with the goal of combining memory and logical functions in
a single device~\cite{wolf,flatte07,flatte,datta,cluj}.
A crucial requirement for most spin devices is the existence of a
robust population of spin-polarized carriers~\cite{zfds} as well
as the feasibility of moving the spins in the device by means of
spin currents. In this context, understanding the various
scattering processes which control the relaxation of the spin and
the spin current has emerged as an important theoretical and
practical problem. In this Letter we focus on the spin current
relaxation  caused by electron-electron scattering in
quasi-two-dimensional quantum wells of finite width.

It is now well established~\cite{cluj,damico,flensberg} that the
Coulomb interaction induces momentum transfer between the carrier
populations of ``up" and ``down" spin. This results in a spin
Coulomb drag (SCD) effect within a single two-dimensional electron
gas (2DEG) layer, in close analogy to the conventional charge
Coulomb drag, which occurs between spatially separated layers.
Because the inherent friction between the two spin components
leads to a decay of the spin current even in the absence of
impurities, the SCD has recently become a subject of  experimental
and theoretical
investigations~\cite{damico2,pustilnik,tse,gros04,weber}.  In
particular, the SCD is known to reduce the spin diffusion constant
relative to the conventional density diffusion constant, and thus
to prolong the time during which a spin packet can be effectively
manipulated.  Indeed a significant reduction in the spin diffusion
constant  was measured by Weber {\it et al.}~\cite{weber} and
found to be in quantitative agreement with the SCD reduction
calculated for a strictly 2DEG in the random phase approximation
(RPA)~\cite{damico2}.

The agreement between the RPA theory for an electron gas of zero
width and the experiment is encouraging, but also somewhat
puzzling. Quantum wells in which the experiments have been
performed are not strictly two-dimensional but have finite
transverse widths of the order of 10~nm.  In contrast to the
conventional charge Coulomb drag, where the inter-layer spacing
causes an exponential suppression of large angle scattering
events, the main contribution to the SCD comes from events with a
momentum transfer of the order of the Fermi momentum. At such
values of the momentum transfer, the form factor that takes into
account the width of the 2DEG is significantly smaller than $1$,
and  should cause a significant {\it reduction} of the SCD  even
at relatively high temperature and densities.

A second problem is that the RPA on which most calculations of
SCD~\cite{damico2,flensberg,tse} so far have been based, neglects
some important physical effects. Namely, in RPA one ignores the
fact that the presence of an electron at a point in space reduces
the probability of other electrons coming nearby, via the
so-called exchange-correlation (xc) hole~\cite{GV05}. The
repercussions of this effect upon the effective electron-electron
interaction are quite subtle because the xc hole, while keeping
the electrons apart, also reduces the collective screening of the
interaction.    Recently we have studied the effect of the  xc
hole in charge Coulomb drag and found that its introduction
results in a significant {\it enhancement} of the charge
transresistivity and qualitative changes in its temperature
dependence~\cite{badalyan}. The sheer size of these effects
suggests that a quantitative description of the SCD cannot be
achieved by a theory that does not include xc effects.

In this Letter we report the results of our calculations of the
SCD  taking into account (i) the finite transverse width of the
quantum well and (ii) xc effects beyond the
RPA~\footnote{Our calculations do not include the spin-orbit
interaction in the one-electron states because the measurement of
Ref.~\cite{weber} were done on a spin grating with wave vector
along the [11] direction of GaAs, which is the direction along
which spin-orbit effects are essentially non existent. We might
add that even in different directions the influence of the
spin-orbit coupling on the spin diffusion constant would show up
as a significant effect only at wave vectors much smaller than the
one studied in Ref.~\cite{weber}. Furthermore, according to a
recent calculation in Ref.~\cite{tse}, the spin Coulomb drag is
enhanced by Rashba spin-orbit coupling, $\gamma$, by a small
factor $1+\gamma^2$, which for the typical value of $\gamma\sim
0.1$, is a $1\%$ correction.},~\footnote{We have neglected the
phonon-mediated SCD. In a single electron layer the
phonon-mediated interaction between electrons cannot compete with
the Coulomb interaction, due to the smallness of the
electron-phonon coupling constant. The phonon-mediated interaction
is effective only in bilayers with large barrier thickness, where
the interlayer Coulomb interaction becomes very small.}. Following
Ref.~\onlinecite{badalyan}, we have used many-body local field
factors~\cite{GV05} to take into account the modifications to the
effective electron-electron interaction~\cite{vs} due to the xc
hole.  The main outcome of our study is that
the enhancement of spin drag caused by the many-body local field
factors  largely compensates the reduction of the effect coming
from the finite well width, thereby restoring good agreement with
the experimental observations of Ref.~\onlinecite{weber}.


SCD manifests itself when a current of active spin-down electrons,
$J_{\downarrow }$, through the Coulomb interaction, transfers
momentum to the passive spin-up electrons with zero current,
$J_{\uparrow}=0$. This process induces a gradient of
electrochemical potential, which results in a  ``spin electric
field" $E_{\uparrow }$. The direct measure of SCD is the spin drag
resistivity, defined as
\begin{equation}
\rho _{\uparrow \downarrow }(T)=E_{\uparrow }/J_{\downarrow
}\text{ when }J_{\uparrow }=0~\text{.}
\end{equation}
We calculate the  spin drag resistivity in the ballistic regime as
appropriate for the clean samples used in the experiments
\cite{weber}. The spin transresistivity is negative. Its absolute 
value at temperature $T$ is given by~\cite{damico}
\begin{eqnarray}
\rho _{\uparrow \downarrow }(T) &=&{\frac{\hbar
^{2}}{2e^{2}n_{\uparrow }n_{\downarrow
}TA}}\sum_{\overrightarrow{q}}q^{2}\int_{0}^{\infty }\frac{d\omega
}{2\pi }\left\vert W_{\uparrow \downarrow }(q,\omega )\right\vert
^{2}  \nonumber \\
&&\times \frac{\text{Im}\Pi _{0 \uparrow}(q,\omega )\text{Im}\Pi
_{0 \downarrow}(q,\omega )}{\sinh ^{2}(\hbar \omega /2T)}~,
\label{eq1}
\end{eqnarray}
where  $\hbar \omega $ and $\hbar \overrightarrow{q}$ are,
respectively, the transferred energy and in-plane momentum from
the spin $\sigma =\uparrow $ to the spin $\sigma =\downarrow $,
$W_{\uparrow \downarrow }(q,\omega )$ is the dynamically screened
effective interaction between electrons with spins $\uparrow $ and
$\downarrow $,  $\Pi _{0 \sigma}(q,\omega )$ is the
spin-$\sigma$  component of the non-interacting finite temperature
polarization function~\cite{GV05}, $n_{\uparrow}$ and $n_{\downarrow }$  are
the densities of up and down spins respectively, and $A$ is the
area of the sample. Further we will assume that the electron gas
has no net spin polarization, i.e. $n_{\uparrow}=n_{\downarrow
}=n/2$.

To calculate the effective interaction
$W_{\uparrow\downarrow}(\omega)$ we have used an approximation
scheme developed a few years ago by  Vignale and Singwi
(VS)~\cite{vs} following the pioneering treatment of Kukkonen and
Overhauser~\cite{kukkonen}. Within this scheme the effective
electron-electron interaction, ${\hat{W}}(q,\omega )$ (a $2\times
2$ symmetric matrix with components $W_{\uparrow\uparrow}$,
$W_{\uparrow\downarrow}$, and $W_{\uparrow\downarrow}$) is
approximated as
\begin{equation}
{\hat{W}}(q,\omega
)={\hat{v}}(q)-{\hat{V}}(q,\omega)\cdot {\hat{\Pi}}(q,\omega
) \cdot {\hat{V}}(q,\omega ) \label{dyson}
\end{equation}
where the full polarization function, ${\hat{\Pi}}(q,\omega )$, is
defined in terms of the non-interacting polarization function,
${\hat{\Pi}}_{0}(q,\omega )$, and the unscreened effective Coulomb
interactions, ${\hat{V}}\left(q,\omega \right) $ (defined below)
has the form
\begin{equation}
\hat\Pi(q,\omega )=\hat \Pi_{0}(q,\omega) \cdot \left[1+\hat
V(q,\omega)\cdot\hat\Pi_{0}(q,\omega )\right]^{-1}~. \label{polar}
\end{equation}
Here all quantities are $2\times 2$ matrices and the dot denotes a
matrix product.  Selecting the $\uparrow\downarrow$ component we
get
\begin{equation}
W_{\uparrow \downarrow }(q,\omega )=V_{\uparrow \downarrow
}(q,\omega )/\varepsilon (q,\omega )+v(q) G_{\uparrow \downarrow
}(q,\omega )F(qd)~, \label{w}
\end{equation}%
where the unscreened effective Coulomb interaction between spins
$\sigma$ and $\sigma ^{\prime }$ is
\begin{equation}
V_{\sigma \sigma ^{\prime }}(q,\omega )=v(q)\left( 1-G_{\sigma
\sigma ^{\prime }}(q,\omega )\right) F(qd)~,  \label{veff}
\end{equation}
where $d$ is the width of the well. Observe how the unscreened
effective interaction $V_{\sigma \sigma ^{\prime }}(q,\omega )$
differs from the bare Coulomb interaction $v(q)=2\pi e^{2}/\kappa
_{0}q$ in two ways.  First, the spin-resolved local field factors
$G_{\sigma \sigma ^{\prime }}$ reduce the bare interaction by a
factor of $1-G_{\sigma \sigma ^{\prime }}(q,\omega )$: this is the
effect of the xc hole, mentioned in the
introduction. Second, the form factor  $F(qd)$ takes into account
the form of the wave function of the lowest transverse subband,
and is defined as the square of the Fourier transform of the
corresponding density profile $\rho(z)$  (for $\rho
(z)=(2/d)\,\sin (\pi z/d)^{2}$ we have $F(qd)=1-(1/3-5/4\pi^2)qd$
for $qd \to 0$ and $F(qd)\to 3/(4\pi^2qd)$ for $qd \to \infty$).
Taking $d=12$ nm and $q\simeq k_F$ we have $q d\simeq 2$ for
$n=4.3\cdot 10^{11}$ cm$^{-2}$ so that $|F(qd)|^2\simeq 0.5$.

In Eq.~(\ref{w}) $\varepsilon (q,\omega )$ is an effective
dielectric function, which can be represented in the following
manner
\[
\varepsilon (q,\omega )=\left( 1+V_{+}(q,\omega )\Pi_{0}(q,\omega
)\right) \left( 1+V_{-}(q,\omega )\Pi_{0}(q,\omega )\right) ~,
\]
where $V_{\pm }(q,\omega )\equiv \left(V_{\uparrow \uparrow }(q,\omega)\pm
V_{\uparrow \downarrow }(q,\omega )\right)/2$.
Introducing the corresponding notation for the local field
factors, $G_{\pm} \equiv (G_{\uparrow\uparrow}\pm
G_{\uparrow\downarrow})/2$, we can write
\begin{eqnarray}
V_{+}(q,\omega ) &=&v(q)\left( 1-G_{+}(q,\omega )\right) F(qd)~,
\label{vpm}
\\
V_{-}(q,\omega ) &=&-v(q)G_{-}(q,\omega )F(qd)~
\end{eqnarray}
$G_+$ and $G_-$ are known as the ``charge-channel" and  the
``spin-channel" local field factors respectively.

In our actual calculations, described below, the local field
factors are derived from  diffusion Monte Carlo
simulations~\cite{senatore,davoudi}) in the low-energy
finite-momentum transfer regime, and from a combination of
analytical and empirical methods~\cite{NCT,qian} in the
small-momentum, high-energy transfer regime. These two regimes
correspond to two distinct mechanism of electron-electron
interaction: in the first, the interaction is mediated by
electron-hole pairs, and in the second by plasmons~\footnote{In
the static regime the local field factors are deduced from the
computed static response of 2DEG to periodic external potentials
of wave vector $q$:  they vary linearly with $q$ both at small $q$
and at  large $q$ but with different slopes as discussed in
Section 5.6 of Ref.~\cite{GV05}.   The local field factor in the
dynamical regime is obtained from a perturbative mode-coupling
theory, described in~\cite{NCT,qian}.}.

Notice that both the dynamic and static local field factors
\cite{qian,davoudi}, which we have used in our calculations, are
evaluated at $T=0$ for an ideal 2DEG of zero width. At present
there are no calculations of the local field factors that take
into account the finite width of the quantum well and/or the
temperature dependence. We expect, however, that the temperature
dependence of the local field factors will play a relatively minor
role in comparison to the temperature dependence that we
explicitly include in the Fermi factors and in the non-interacting
polarization functions. Similarly, the effect of the finite width
of the well should be largely taken care of by our use of the form
factors $F(qd)$ in Eqs.~(\ref{w}) and (\ref{veff}).


\begin{figure}[t]
\centering\includegraphics[width=0.56\linewidth,angle=-90]{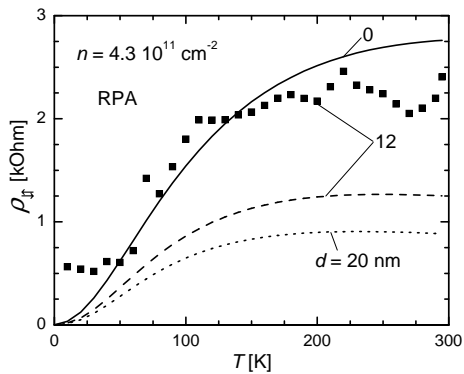}
\caption{Spin drag resistivity $\rho_{\uparrow\downarrow}$ as a 
function of temperature, calculated within RPA. The solid, dashed, 
and dotted curves correspond, respectively, to 
$\rho_{\uparrow\downarrow}$ for $d=0$, $12$, and $20$ nm. 
The black squares represent $\rho_{\uparrow\downarrow}$, deduced 
from the experimental data reported in Fig.~3 of 
Ref.~\onlinecite{weber} for the sample with 
$n=4.3\cdot 10^{11}$ cm$^{-2}$ and $d=12$ nm.} \label{fig1}
\end{figure}
Recently Weber \textit{et al.} \cite{weber} have demonstrated
experimentally that the spin diffusion constant of electrons in an
n-type GaAs quantum well is strongly reduced by the
Coulomb interaction over a broad range of temperatures and
electron densities. Their measurements were  in good
agreement with the theoretical calculations of
D'Amico and Vignale \cite{damico2}, which, however,
neglected both the  finite width of the quantum well and the
local field factor. Here we present
results of calculations which include both these effects.

First we show that the effect of the finite width of the quantum
well is very important. In Fig.~\ref{fig1} we plot the absolute
value of the spin drag resistivity, calculated within RPA, vs
temperature for three different values of the width, $d=0,~12$,
and $20$ nm. The solid curve reproduces the RPA calculations of
the spin drag resistivity for an ideal 2DEG of zero
width~\cite{damico2}. The black squares  show the spin drag
resistivity, as deduced from the experimentally measured spin
diffusion constant $D_s$ according to the formula
\cite{damico3,weber}
\begin{equation}
\frac{D_{c}}{D_{s}}=\frac{\rho}{\rho+\rho_{\uparrow \downarrow}}
\frac{\Pi_{s}}{\Pi_{0}}~. \label{diffusion}
\end{equation}
Here the charge diffusion constant $D_{c}$ is obtained from the
electric resistivity $\rho$ via the Einstein relation, and
$\Pi_s/\Pi_0$ is the many-body enhancement of the spin
susceptibility of the electron gas, which, for the time being,
following Ref.~\onlinecite{weber}, we approximate as $1$.
Fig.~\ref{fig1} shows that an increase in the width of the quantum
well is accompanied by a strong reduction of the spin drag
resistivity: at $T=T_{F}$ ($T_{F} \simeq 178$ K) $\rho _{\uparrow
\downarrow }$ for $d=12$ nm is a factor of two smaller than at
$d=0$. The physical reason for this effect is that the SCD is
dominated by large angle scattering events, where the momentum
transfer $q$ is of the order of the diameter of the Fermi sphere.
At these values of $q$ the form factor $F(qd)$ is substantially
smaller than $1$ -- the more so the higher the density. By
contrast, in the ordinary charge Coulomb drag large angle
scattering events are exponentially suppressed for $q>1/\Lambda$
where $\Lambda$ is the inter-layer separation.  Then, the finite
width of the quantum well produces corrections of the order of
$|F(d/\Lambda)|^2$, which is rather close to $1$.

\begin{figure}[t]
\centering\includegraphics[width=0.56\linewidth,angle=-90]{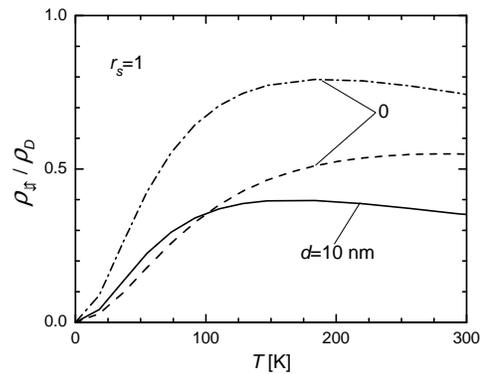}
\caption{The scaled spin drag resistivity 
$\rho_{\uparrow\downarrow}/\rho_D$ vs $T$ for $r_s=1$. The solid 
and dash-dotted curves are calculated for a 2DEG of the width 
$d=0$ and $10$ nm within the VS scheme. The dashed curve corresponds 
to $\rho_{\uparrow\downarrow}/\rho_D$ of an ideal 2DEG, calculated 
within the RPA.
The ordinary Drude resistivity, $\rho_D$, is calculated for
the mobility $\mu =300$ V cm$^{-1}$ s$^{-1}$.} \label{fig2}
\end{figure}
Next we include the effect of exchange and correlation on the spin
drag resistivity. In Fig.~\ref{fig2} we compare the spin drag
resistivity, calculated within the VS scheme, with the
corresponding result obtained within the RPA. The scaled spin drag
resistivity is shown as a function of $T$ for the dimensionless 
parameter $r_{s}=1$
corresponding to $n=3.5\cdot 10^{11}$ cm$^{-2}$ and to the Fermi
temperature $T_F=147$ K.  Notice that the inclusion of the local
field factors shifts the peak of the spin drag resistivity towards
lower temperatures compared to the RPA. Accordingly,  the spin
drag resistivity curves show quite different slopes at high
temperatures depending on whether local field factors are included
or not.
As a result, the reduction in spin drag coming from the finite 
width of the well is largely compensated by the many-body 
enhancement of the spin drag
resistivity. The situation is similar for $r_{s}=2$. We conclude
that the combined effect of the finite width and the
xc hole on the effective interaction restores
the agreement between theory and experiment.
\begin{figure}[t]
\centering\includegraphics[width=0.56\linewidth,angle=-90]{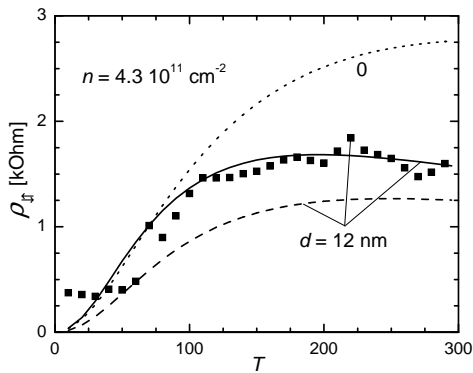}
\caption{Spin drag resistivity $\rho_{\uparrow\downarrow}$ as a 
function of temperature for
the electron density $n=4.3\cdot 10^{11}$ cm$^{-2}$. The dashed
and dotted curves correspond to $\rho_{\uparrow\downarrow}$,
calculated within the RPA for $d=0$ and $12$ nm, respectively. 
The solid curve correspond to $\rho_{\uparrow\downarrow}$, 
calculated beyond the RPA within the VS scheme for $d=12$ nm. 
The symbols represent $\rho_{\uparrow\downarrow}$, deduced from 
the experimental data for $d=12$ nm of Ref.~\onlinecite{weber}. 
Notice that the data points for the spin drag resistivity are lower 
than in Fig. 1, because we have taken into account the many-body 
enhancement of the spin susceptibility in converting the experimental 
values of $D_s$ to $\rho_{\uparrow\downarrow}$ according to 
Eq.~(\ref{diffusion}).}
\label{fig3}
\end{figure}

In Fig.~\ref{fig3} we verify this conclusion for the strict
experimental situation of Ref.~\onlinecite{weber}. Here the dotted
and dashed curves are calculated within RPA, respectively, for
$d=0$ and $12$ nm and show the strong reduction of spin drag due
to the form factor of the quantum well. The spin drag resistivity
is deduced from the experimental data for the spin diffusion via
Eq.~(\ref{diffusion}). However, we have now taken into account the
many-body enhancement of the spin susceptibility with respect to
the non-interacting spin susceptibility. In a paramagnetic
electron liquid the Hartree terms cancel and the enhancement is
determined by the ``spin-channel" many-body local field factor
$G_{-}(q,0) $ according to the formula $\Pi_{0}/\Pi_{s}=1+\lim_{q
\to 0} V_{-}(q,0)\Pi_{0}(q,0)$~\cite{GV05}. Then, it is seen that
our theoretical curve for the spin drag resistivity, including
both the many-body effects beyond RPA and the finite width of the
quantum well, agrees very well with the experimental data of
Ref.~\cite{weber}, while the RPA, in the presence of the form
factor, does not.

We acknowledge support from the Volkswagen Foundation and NSF
Grant No. DMR-0313681. SMB thanks J. Fabian for useful discussions
and acknowledge additional support from Sonderforschungsbereich
689. We thank C. P. Weber for kindly sending us their experimental
data.


\begin{thebibliography}{99}
\bibitem{wolf} S. A. Wolf \textit{et al.}, Science \textbf{294}, 
1488 (2001) and references therein.

\bibitem{flatte07}D. D. Awschalom and M. E. Flatt\'e, Nature Physics
 {\bf 3}, 153 (2007).

\bibitem{flatte} M. E. Flatt\'e and G. Vignale, J. App. Phys. \textbf{78},
1273 (2001).

\bibitem{datta} S. Datta, and B. Das, Appl. Phys. Lett. \textbf{56}, 665
(1990).

\bibitem{cluj} G. Vignale, in {\it Manipulating Quantum Coherence in
Solid State Systems},  edited by M. E. Flatt\'e and I. Tifrea, Springer, 2007.

\bibitem{zfds} I Zutic, J. Fabian, and S. Das Sarma, Rev. Mod. Phys.
\textbf{76}, 323 (2004).

\bibitem{weber} C. P. Weber, N. Gedik, J. E. Moore, J. Orenstein,
J. Stephens, and D. D. Awschalom, Nature, \textbf{437}, 1330
(2005).

\bibitem{damico} I. D'Amico and G. Vignale, Phys. Rev. B
\textbf{62}, 4853 (2000).

\bibitem{flensberg} K. Flensberg, T. S. Jensen, and N. A. Mortensen,
Phys. Rev. B \textbf{64}, 245308 (2001).

\bibitem{damico2} I. D'Amico and G. Vignale, Phys. Rev. B \textbf{68},
045307 (2003).

\bibitem{pustilnik} M. Pustilnik, E. G. Mishchenko, and O. A. Starykh,
Phys. Rev. Lett. \textbf{97}, 246803 (2006).

\bibitem{tse} W.-K. Tse and S. Das Sarma, Phys. Rev. B \textbf{75},
045333 (2007).

\bibitem{gros04}K. Luis and C. Gros, New Journal of Physics  {\bf 6}, 187 (2004).

\bibitem{badalyan} S. M. Badalyan, C. S. Kim, G. Vignale, and G. Senatore,
Phys. Rev. B \textbf{75}, 125321 (2007).

\bibitem{GV05} \textit{Quantum Theory of the Electron Liquid}, G. F.
Giuliani and G. Vignale, (Cambridge University Press, Cambridge,
2005).

\bibitem{vs} G. Vignale and K. S. Singwi, Phys. Rev. B \textbf{31}, 2729
(1985); \textit{ibid} \textbf{32}, 2156 (1985).


\bibitem{kukkonen} C. A. Kukkonen and A. W. Overhauser, Phys. Rev. B
\textbf{20}, 550 (1979).

\bibitem{senatore} G. Senatore, S. Moroni, and D. M. Ceperley, in
\textit{Quantum Monte Carlo Methods in Physics and Chemistry},
edited by M. P. Nightingale and C. J. Umrigar, Kluver, Dordrecht,
1999.

\bibitem{davoudi} B. Davoudi, M. Polini, G. F. Giuliani, and M. P. Tosi,
Phys. Rev. B \textbf{64}, 153101 (2001); \textit{ibid}
\textbf{64}, 233110 (2001).

\bibitem{NCT} R. Nifos\'i, S. Conti, and M. P. Tosi, Phys. Rev. B \textbf{58},
12758 (1998).

\bibitem{qian} Z. Qian and G. Vignale, Phys. Rev. B 65, 235121
(2002); Z. Qian, {\it ibid} B 72, 075115 (2005).

\bibitem{damico3} I. D'Amico and G. Vignale, Europhys. Lett. \textbf{55}, 566
(2001).





\end{thebibliography}
\end{document}